\documentclass[useAMS,usenatbib]{mn2e}

\usepackage[activeacute,swedish,english]{babel}
\usepackage{graphicx}
\usepackage[T1]{fontenc} 
\usepackage{latexsym}
\usepackage{verbatim}
\usepackage{amsmath}
\usepackage{subfigure}

\def\gtappeq{\mathrel{ \rlap{\raise.5ex\hbox{$>$}}
                      {\lower.5ex\hbox{$\sim$}}  } }

\def\leappeq{\mathrel{ \rlap{\raise.5ex\hbox{$<$}}
                      {\lower.5ex\hbox{$\sim$}}  } }                      

\title[J1507: An eclipsing period bouncer in the Galactic halo]{The cataclysmic variable SDSS J1507+52: An eclipsing period bouncer in the Galactic halo} 

\author[H. Uthas et al.]{Helena Uthas$^{1}$\thanks{E-mail:
h.uthas@astro.soton.ac.uk}, Christian Knigge  $^{1}$, Knox S. Long$^{2}$, Joseph Patterson,$^{3}$   \newauthor and John Thorstensen$^{4}$ \\\
$^{1}$University of Southampton, Department of Physics and Astronomy,
  Highfield, Southampton SO17 1BJ, UK\\ 
$^{2}$Space Telescope Science Institute, 3700 San Martin Drive, Baltimore, MD 21218, USA\\
$^{3}$Columbia University, Department of Astronomy, New York, NY, USA\\
$^{4}$Department of Physics and Astronomy, Dartmouth College, Hanover, NH, USA}

\begin{document}

\date{Accepted year month date. Received year month date; in original
form  year month date} 

\pagerange{\pageref{firstpage}--\pageref{lastpage}} \pubyear{year}

\maketitle

\label{firstpage}

\begin{abstract}

SDSS J1507+52 is an eclipsing cataclysmic variable consisting of a cool, non-radially pulsating white dwarf and an unusually small sub-stellar secondary. The system has a high space velocity and a very short orbital period of about 67 minutes, well below the usual minimum period for CVs. To explain the existence of this peculiar system, two theories have been proposed. One suggests that SDSS J1507+52 was formed from a detached white-dwarf/brown-dwarf binary. The other theory proposes that the system is a member of the Galactic halo-population. 

Here, we present ultraviolet spectroscopy of SDSS J1507+52 obtained with the Hubble Space Telescope with the aim of distinguishing between these two theories. The UV flux of the system is dominated by emission from the accreting white dwarf. Fits to model stellar atmospheres yield physical parameter estimates of $T_{\text{eff}} = 14200 \pm 500$ K, $\log g = 8.2\,\pm\,0.3$, $v \sin i = 180 \pm 20$ km\,s$^{-1}$ and $[Fe/H] = -1.2 \pm 0.2$. These fits suggest a distance towards SDSS J1507+52 of $d = 250 \pm$ 50 pc. The quoted uncertainties include systematic errors associated with the adopted fitting windows and interstellar reddening.

Assuming that there is no contribution to the UV flux from a hot, optically thick boundary layer, we find a $T_{\text{eff}}$ much higher than previously estimated from eclipses analysis. The strongly sub-solar metallicity we infer for SDSS J1507+52 is consistent with that of halo stars at the same space velocity. We therefore conclude that SDSS J1507+52 is a member of the Galactic halo.
 

\end{abstract}

\begin{keywords}
binaries: close $-$ stars: dwarf novae $-$ stars: individual: SDSS J1507+52 $-$ novae, cataclysmic variables.
\end{keywords}

\section{Introduction} \label{intro}

Cataclysmic Variables (CVs) are binary systems where a late-type main-sequence donor star transfers mass onto a primary white dwarf (WD). According to standard evolutionary theory, loss of angular momentum drives CVs to initially evolve from longer to shorter orbital periods. All systems are predicted to reach a minimum orbital period roughly when the donor stops its hydrogen burning and becomes degenerate. The observed minimum period is found at $\approx$ 83 minutes~\cite{2009MNRAS.397.2170G}. At about the point when the donor reaches this evolutionary state, the thermal time scale becomes longer than the mass-transfer timescale, and the donor cannot shrink fast enough in response to continued mass loss. As a result, the brown-dwarf donor will expand, and the system will move towards longer orbital periods. 

Standard evolutionary theory also predicts that about 70\% of all CVs should have passed their minimum period and have sub-stellar donors (\citealp{1993A&A...271..149K}; \citealp{1997MNRAS.287..929H}). However, few such systems have been successfully confirmed, and until recently, almost no CVs containing donors with masses below the hydrogen-burning limit, had been found. Systems that have passed their minimum period (post-bounce systems) should be very faint and are therefore difficult to find, in general. Nevertheless, in the course of the Sloan Digital Sky Survey (SDSS), a few post-bounce CVs were discovered (\citealp{2008MNRAS.388.1582L}), among them the eclipsing system SDSS J1507+52 (hereafter J1507). 

This particular system was quickly recognised to be odd due to its short orbital period of about 67 minutes (\citealp{2005AJ....129.2386S}), which is well below the minimum period characteristic of normal CVs. A few other systems are occasionally found to have periods below the usual minimum period, but, in general, these systems show evidence for abnormally hot and bright donor stars in their optical spectra, suggesting that their secondaries are nuclear-evolved objects. This is not the case in J1507, whose secondary is not visible at all in optical spectroscopy (\citealp{2005AJ....129.2386S}; \citealp{2007MNRAS.381..827L}). Together with its short orbital period, this indicates a system with a faint disc, a low accretion rate and a relatively unevolved, low-mass donor.

\cite{2007MNRAS.381..827L} performed eclipse analysis of J1507 and found a donor mass of 0.056 $\pm$ 0.001 M$_{\odot}$, clearly below the hydrogen-burning limit. However, because of its anomalously short orbital period, the system is not consistent with the standard mass-period relation for CVs (\citealp{2006MNRAS.373..484K}). In line with this,~\cite{2007MNRAS.381..827L} found the donor radius to be smaller than predicted by standard CV evolution sequences. They therefore suggested that the donor might be unusually young, i.e. that J1507 might be a CV that formed recently from a previously detached WD -- brown dwarf close binary system. A young brown dwarf has a higher density, and therefore a smaller radius, since the donor has not yet had a chance to expand in response to mass loss. This would explain the short orbital period found in J1507, since according to the period-density relation, a higher density and subsequently a smaller radius, implies a shorter orbital period. The list of photometrically inferred system parameters provided by~\cite{2007MNRAS.381..827L} also includes the effective temperature of the white dwarf, which they estimate to be $T_{\text{eff}}$ = 11000 $\pm$ 500 K. 

Concurrently,~\cite{2008PASP..120..510P} showed that the system has an unusually high space velocity, similar to the velocities of stars in the Galactic halo. A typical star of 1 M$_{\odot}$ in the Galactic disc has a space velocity below 50 km\,s$^{-1}$, while J1507 has a velocity of about 167 km\,s$^{-1}$. If J1507 is a member of the Galactic halo, its donor will be a Population II object with low metallicity. Due to their lower atmospheric opacity, such objects have significantly smaller radii than their solar metallicity Population I counterparts. Membership of the Galactic halo would therefore also account for J1507's small donor radius and short orbital period. However, Population II stars only represent about 0.5 \% of the stars in the Solar neighbourhood, which would make J1507 a rare system.~\cite{2008PASP..120..510P} estimated a slightly higher effective temperature of the WD in J1507, $T_{\text{eff}}$ = 11500 $\pm$ 700 K. They also found the system to exhibit multi-periodic variability on time-scales of minutes, which they interpreted as non-radial pulsations originating from the primary white dwarf.

Both hypothesis imply that J1507 is an interesting and important system that can shed light on poorly understood aspects of CV evolution. However, since the models are very different, it is clearly important to determine which -- if either -- of them is correct. As noted above, eclipse-based estimates for the effective temperature of the WD suggested $T_{\text{eff}} \simeq$ 10000 K -- 12000 K. At these temperatures, the ultraviolet (UV) spectrum of a WD accreting metal-rich material would be expected to show strong absorption features due to Fe II and Fe III (c.f. \citealp{1994ApJ...426..294H}). We have therefore obtained far- and near-UV spectroscopy of J1507 with COS and STIS onboard the Hubble Space Telescope, with the aim of measuring the metallicity of the system and establishing its status as either a Pop I (disk) or a Pop II (halo) object.


\begin{figure}
\includegraphics[width=9cm]{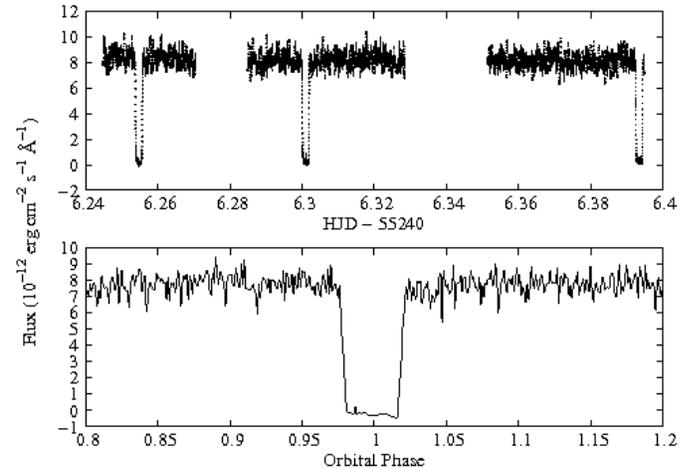}
\caption{The Gaussian smoothed and flux-calibrated light curve obtained by COS/HST, constructed from the monochromatic flux averaged over the region 1425\,\AA\, -- 1900\,\AA. Bottom panel shows the phase-folded light curve in the vicinity of the eclipse.} 
\label{fig:light}
\end{figure}


\section{Observations and Reduction} 

Far- and near-UV observations of J1507 were performed in February 2010, using the HST. The observations were obtained in the so-called time-tag mode. In this mode, each photon event is tagged with a corresponding event time, along with their wavelength and spatial position on the detector. This allows us to rebin the data over any time interval and wavelength region. 

Far-UV observations were carried out with the Cosmic Origins Spectrograph (COS), for a total exposure time of about 3 hours, covering 3 eclipses in three successive HST orbits. The maximum time resolution of the COS TIME-TAG mode is 32 ms. The G140L grating was used, which gave a spectral range of 1230\,\AA\ -- 2378\,\AA\,and a spectral resolution of 0.5\,\AA.  

Near-UV observations were obtained by the Space Telescope Imaging Spectrograph (STIS), for a total exposure time of about 1.4 hours, at a maximum time resolution 125 $\mu$s. The G230L grating was used, resulting in a spectral range of 1650\,\AA\ -- 3150\,\AA\,and a spectral resolution of 3.16\,\AA. The data set consists of 3 sub-exposures (of length, 1600 s, 1918 s and 1708 s). No near-UV data was obtained during the eclipses. Also, gaps in the data are cased by interruptions coming from Earth occultation. 

All data were reduced and calibrated using the \textsc{Pyraf} package \textsc{Stsdas}, provided by STScI (the Space Telescope Science Institute)\footnote{\textsc{Stsdas} and \textsc{Pyraf} are products of the Space Telescope Science Institute, which is operated by AURA for NASA.}.


\begin{figure*}
\includegraphics[width=11.0cm]{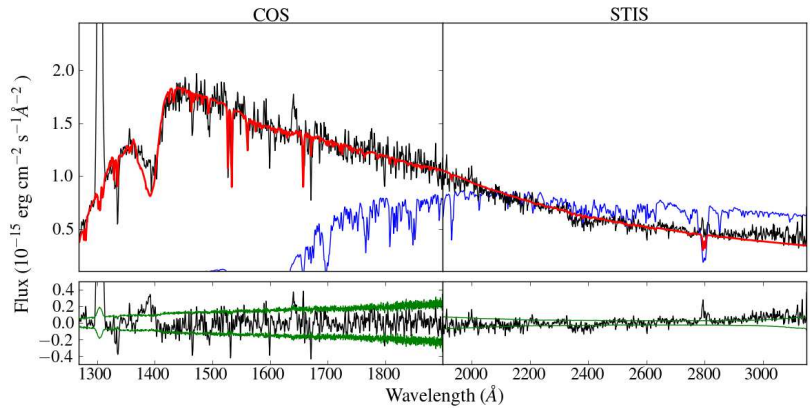}
\caption{The two top panels show the mean out-of-eclipse spectrum for both the far (left) and near (right) UV regions. A model constructed from parameters obtained from the far UV-region is plotted in red (see Table~\ref{pars}). A model corresponding to values from Littlefair et al. (2007), $T_{\text{eff}}$=11000 K and $\log g$=8.5, is plotted in blue. Underneath each wavelength region is a plot showing the residuals. The dips at (1460, 1530, 1600 and 1670)\,\AA\, coincide with regions of bad pixels. The 1-$\sigma$ flux errors are plotted in green.} 
\label{fig:spec}
\end{figure*}


\section{Analysis and Results}

Figure~\ref{fig:light} shows the Gaussian smoothed and flux-calibrated light curve from the far-UV data. This light curve was constructed by dividing the larger exposures into one-second bins, and by average the fluxes over the wavelength range, 1425\,\AA\, -- 1900\,\AA. The light curve shows sharp, square-shaped eclipses, close to zero flux at mid-eclipse, and a very flat out-of-eclipse continuum, strongly suggesting that the UV flux is dominated by emission from (or very close to) the accreting WD. The folded light curve constructed from the combined data for all three HST orbits is presented in the bottom panel. All further analysis was carried out on the unsmoothed data, excluding the eclipses. Ground-based optical observations were performed at the same time as the HST observations, yielding a magnitude for J1507 of $g = 18.44 \pm 0.02$. This is consistent with SDSS, showing that the system was in quiescence during the HST observations. 

The two top panels of Figure~\ref{fig:spec} show the mean out-of eclipse spectrum (in black), for both the near- and far-UV. The reduction was performed separately for the COS and STIS data, and the mean spectrum for each region were joined at 1900 \AA\,without overlap for a total range of 1268\,\AA\, -- 3150\,\AA. The COS data was rebinned slightly, to 0.2\,\AA/pix, to increase the S/N and provide sufficient sampling of the G140L grating's spectral resolution. Different dispersions were retained for the the far- and near-UV regions (0.2 \,\AA/pix and 1.58 \,\AA/pix, respectively). The overall UV spectrum for J1507 is relatively sparse in spectral lines. We identify CII at 1335, SiII at 1526.7 HeII at 1640.5, AlII at 1671 and MgII at 2800 \,\AA. The strong emission feature near 1300 \AA\, is almost certainly due to geocoronal emission in the O I 1304 \AA\, line that could not be cleanly removed by the pipeline background subtraction.

\subsection{Spectral Modelling}

A model grid spanning the four key atmospheric parameters, effective temperature ($T_{\text{eff}}$), surface gravity ($\log g$), metallicity ($[Fe/H]$) and rotational velocity ($v \sin i$), was constructed in order to fit the far- and near-UV spectrum. Atmospheric structures were calculated with \textsc{Tlusty}, with the spectral synthesis being done with \textsc{Synspec} (\citealp{1988CoPhC..52..103H}; \citealp{1995ApJ...439..875H}). The grid consisted of; 12500 K $\leappeq T_{\text{eff}} \leappeq$ 20000 K in steps of 500 K, 7.5  $\leappeq \log g \leappeq 9.5$ in steps of 0.25, -2.0 $\leappeq [Fe/H]  \leappeq$ 0.0 in steps of 0.25, and 0 km\,s$^{-1} \leappeq v \sin i \leappeq$ 1000 km\,s$^{-1}$ in steps of 50 km\,s$^{-1}$. Models at intermediate parameter values were constructed by linear interpolation on this grid. Pixels with non-zero data quality flags set were excluded from our fits, as well as the regions around the spectral lines, OI at 1304\,\AA\, CII at 1335\,\AA\,and the quasi-molecular H feature at $\sim$ 1400\,\AA. 

Before our models could be compared to the observed spectrum, they were convolved with a Gaussian filter to the spectral resolution appropriate to the COS (< 1900\,\AA) and STIS (> 1900\,\AA) data, and linearly interpolated onto the observational wavelength scale. In general, our model fits did represent the main features of the data quite well, but the formal $\chi^{2}$ was somewhat high. In order to obtain more realistic formal parameter errors, we therefore added an intrinsic dispersion term to the flux errors in such a way that $\chi^{2}_{\nu}$ = 1. This dispersion corresponds to about 1\,\% of the flux at 1900\,\AA. In order to explore the systematic uncertainties affecting our fits, we tried fitting many different wavelength regions, such as the far-UV (COS), near-UV (STIS), regions with a high line concentration and also the whole wavelength range spanning both the far- and near-UV.

\begin{figure*}
\includegraphics[width=11.0cm]{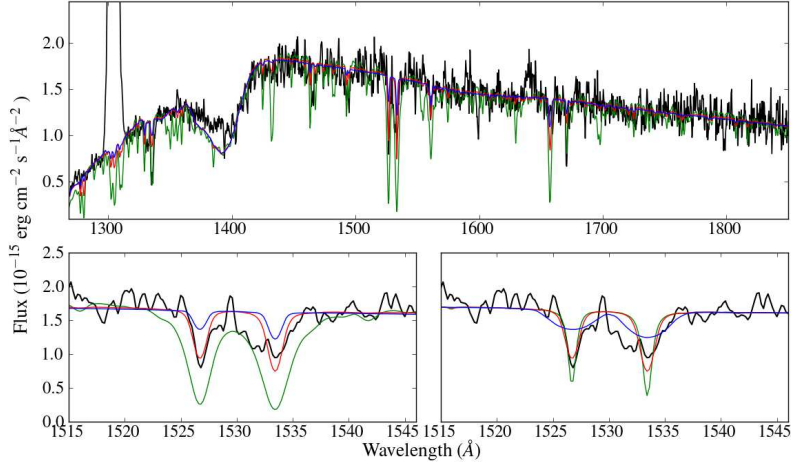}
\caption{The top panel shows the mean out-of-eclipse spectrum for the far UV region, where models at three different $[Fe/H]$, 0 (green), -1.2 (red) and -2 (blue), are plotted on top. The bottom left panel shows a  zoom of the region around the Si II, where the coloured lines correspond to the same values of the metallicity as in the top panel. In the bottom right panel, the colours correspond to $v \sin i$ at 0 km\,$s^{-1}$ (green), 180 km\,$s^{-1}$ (red), 500 km\,$s^{-1}$ (blue).} 
\label{fig:met}
\end{figure*}

\subsection{Temperature and Metallicity}

A fit to the far UV-region 1268\,\AA\, -- 1900\,\AA\,\,yields $T_{\text{eff}} = 14200 \pm 50$ K, $\log g = 8.2 \pm 0.04, [Fe/H] = -1.2 \pm 0.05 $ and $v \sin i = 180 \pm 20$ km\,s$^{-1}$. These fit parameters provide a good approximation also for the near-UV spectrum out to 2800\,\AA\, where the spectrum may begin to show disc features. We adopt these parameters as our best-fit parameters. The quoted errors are formal errors from the $\chi^{2}$ fitting. Table~\ref{pars} presents the best parameters along with their uncertainties. These uncertainties are larger than the formal errors, because they include our best estimates of the systematic uncertainties associated with different choices of fitting windows. However, there are several reasons for adopting the parameters inferred from the fit to the COS far-UV data set as our best-bet estimates. This region includes almost all line features, and is also most sensitive to changes in $T_{\text{eff}}$ and $\log g$. Also, the far UV-region is more likely to represent pure light from the WD, while towards redder wavelengths, the spectrum could be affected by the disc and the bright spot. Finally, fitting only the COS data removes any possibility that a mis-match in the flux calibration of COS and STIS could affect our results.

The two top panels in Figure~\ref{fig:spec} show the best-fit model in red, along with the mean out-of-eclipse spectrum for J1507. The corresponding residuals between model and data are plotted underneath each of the COS and STIS spectral regions, together with the 1-$\sigma$ error range marked in green. The dips in the residuals at (1460, 1530, 1600 and 1670)\,\AA\,, are caused by bad pixels in the data and these regions were excluded from our fits. As a comparison, a model showing the parameters by~\cite{2007MNRAS.381..827L}, $T_{\text{eff}}$ = 11000 K and $\log g$ = 8.5, is plotted in blue. Clearly, this temperature is far too low to match the data.
 
 The top panel in Figure~\ref{fig:met}, shows the mean spectrum plotted together with models at three different values of $[Fe/H]$, 0 (green), -1.2 (red) and -2 (blue). It is clear that a model of solar metallicity (green line) does not match the data and overestimates both the number and strength of visible absorption lines in the spectrum. The left panel shows a zoom of the region around the Si II line (the coloured lines correspond to the same values of the metallicity as given for the top panel). To the right is a plot showing the dependence of the model on $v \sin i$, where the coloured lines in green, red and blue represent rotational velocities of 0 km\,s$^{-1}$, 180 km\,s$^{-1}$ and 500 km\,s$^{-1}$, respectively. In all panels, the red lines show the best fitted parameters. The metallicity and rotational velocity affect both the line depths and widths. Therefore, as a check, we fixed $T_{\text{eff}}$ and $\log g$ to our best values, and made model fits to smaller wavelength regions specifically chosen because they contain strong and/or many spectral lines, with $v \sin i$ and $[Fe/H]$ as free parameters. We find consistent values of $v \sin i$ and $[Fe/H]$, independent of the wavelength region we perform the fits on.   
  
We compared the metallicity found for J1507 with the metallicity distribution for Galactic halo stars at the same space velocity. ~\cite{2008PASP..120..510P} calculated a space velocity for J1507 of 167 km\,s$^{-1}$, which can be broken into the Galactic velocity components U, V and W, where $\sqrt{U^{2} + W^{2}}$ =139 km\,s$^{-1}$ for J1507.~\cite{1996AJ....112..668C} presented a study of the metallicity distribution for the Galactic halo stars. They find that for stars with velocities in the range 100 <  $\sqrt{U^{2} + W^{2}}$ < 140, the metallicity distribution peaks around -0.7, but has a long tail towards metallicities as low as -1.5 (see Figure 6 in their paper). This is in good agreement with our metallicity of -1.2. Therefore, we conclude that J1507 is most likely a Galactic halo CV.

\subsection{Possible Extinction Effects}

Up to now, we have assumed that the observed far-UV spectrum of SDSS J1507 is unaffected by interstellar extinction. Based on the absence of the well-known 2175\,\AA\,absorption feature in the data, we can set E(B-V) $\leq$ 0.05 as a fairly conservative limit on the amount of reddening that may be present. In order to test the effect of extinction at this level on our conclusions, we carried out additional model fits after dereddening the data by E(B-V) = 0.05. We find that none of our inferred parameters would change significantly if extinction at this level were present.



\subsection{Distance Estimates}

The theoretical relationship between observed flux ($F$) and the Eddington flux ($H$) provided by the \textsc{Synspec} models can be used to estimate the distance ($d$) towards J1507. More specifically, $F =  4 \pi R_{\text{wd}}^{2}H/\,d^{2}$, where $R_{\text{wd}}$ is the WD radius. This means that the normalisation factor needed to optimally match a model spectrum to the data is a direct measure of $R_{\text{wd}}^{2}/d^{2}$. WDs also follow the well-known mass-radius relation (with only a weak temperature sensitivity), so the surface gravity of the model, $g = G\,M_{\text{wd}}/R_{\text{wd}}^{2}$, uniquely fixes $R_{\text{wd}}$. Thus for a WD model with given $T_{\text{eff}}$ and $\log g$, the normalisation factor required to fit the data is a unique measure of distance. 

In practice, we estimated the distance by first fitting a linear function to the relationship between $M_{\text{wd}}$ and $\log g$ in Pierre Bergeron's WD cooling models\footnote{Cooling models by Pierre Bergeron are found at: \newline http://www.astro.umontreal.ca/$\sim$bergeron/CoolingModels/}, at $T_{\text{eff}}$ = 14500 K. We then use this function to estimate $M_{\text{wd}}$, and hence $R_{\text{wd}}$, for given $\log g$. Rough values of the WD mass and radius are found at $M_{\text{wd}} = 0.75 \pm 0.15$ M$_{\odot}$ and $R_{\text{wd}} = 0.011 \pm 0.002$ R$_{\odot}$. This is consistent with both ~\cite{2007MNRAS.381..827L} and ~\cite{2008PASP..120..510P}. The normalisation constant of the model combined with $R_{\text{wd}}$, yield the corresponding distance estimate. We find a distance towards J1507 of $d = 250 \pm 50$ pc (the effect of reddening is allowed for in the quoted error).


\begin{table}
  \centering
  \begin{minipage}{70mm}
  \caption{Best-fit parameters for J1507 obtained from model fitting to the far-UV region. The errors are defined as the whole range for which we find good fits, irrespective of wavelength region. Formal 1-$\sigma$ errors are given in parenthesis.}
 \begin{tabular}{ll}
 \hline
 \hline
$T_{\text{eff}}$: & 14200 $\pm$ 500 (50) K  \\
$\log g$: & 8.2 $\pm$ 0.3  (0.04)\\
$[Fe/H]$: & -1.2 $\pm$ 0.2 (0.05)   \\
$v \sin i$: & 180 $\pm$  20 (20) km\,s$^{-1}$\\
\hline
\hline
\label{pars} 
\end{tabular}
\end{minipage}
\end{table}


\section{Discussion and Summary}

We have obtained HST observations  in the UV spectral-range of the cataclysmic variable J1507, with the aim of measuring its metallicity to determine whether or not the system is a member of the Galactic halo. 

By comparing the observed spectrum to synthetic spectra described by the four parameters, $T_{\text{eff}}$, $\log g$, $[Fe/H]$ and $v \sin i$, a best fit is found at $T_{\text{eff}}$ = $14200 \pm 500$ K. This value of $T_{\text{eff}}$ is based on the assumption that the WD is the only component contributing to the UV flux in J1507. If a hot, optically thick boundary layer would be present, it would bias our estimate of $T_{\text{eff}}$. However, the single-temperature WD models presented here seem to provide a good fit to the data, and we expect that the boundary layer in a low $\dot{M}$-system (such as J1507) would be optically thin and thus not contribute significantly to the UV flux. 

Our best model fit give a $T_{\text{eff}}$ that is much higher than previous estimates, which implies that the accretion rate is higher than previously suggested. More specifically, since $\dot{M} \propto T^{4}$, the 30\% increase in the estimated  $T_{\text{eff}}$ corresponds to an increase in $\dot{M}$ by almost a factor of 3 (\citealp{2003ApJ...596L.227T}). This may have significant implications for the evolution of this system, including the question of whether gravitational radiation alone is sufficient to drive this accretion rate (see \citealp{2009ApJ...693.1007T}; \citealp{2011arXiv1102.2440K}).

At the higher $T_{\text{eff}}$ we infer, Fe II and III are no longer dominant contributors to the atmospheric UV opacity, making it more difficult than expected to measure the metallicity of the system. Nevertheless, model fits to the data clearly favour a significantly sub-solar metallicity, $[Fe/H] = -1.2 \pm 0.2$, comparable to the typical metallicity found for halo stars at the same high space velocity as J1507.

\cite{2008PASP..120..510P} found pulsations in J1507, which they identified as non-radial pulsations originating from the primary WD. Non-accreting CVs show pulsations within a well-defined temperature range, the so-called, instability strip at 10900 K -- 12200 K (\citealp{2006AJ....132..831G}). We find that the effective temperature in J1507 is well above the instability strip for non-accreting pulsating WDs. However, it is already becoming clear that non-radially pulsating WDs in CVs are found across a wider temperature range, often towards higher temperatures (\citealt{2010ApJ...710...64S}). The amplitudes of these pulsations are expected to vary with wavelength, and a search for the UV counterpart of the optical WD pulsations will be presented in Uthas et al. (in preparation). The fact that non-radial pulsations are present in J1507 is certainly interesting and potentially important since a higher metallicity in the outer envelope of the accretors might be able to push the instability strip towards higher temperatures (\citealt{2006ApJ...643L.119A}). In this context, it is interesting to note that J1507 has a very low metallicity, but is nevertheless managing to pulsate at an effective temperature of above 14000 K.
   
\bibliography{ref}

\begin{thebibliography}{18}
\expandafter\ifx\csname natexlab\endcsname\relax\def\natexlab#1{#1}\fi

\bibitem[{Arras} et~al.(2006){Arras}, {Townsley} \&
  {Bildsten}]{2006ApJ...643L.119A}
{Arras} P., {Townsley} D.~M., {Bildsten} L., 2006, \apjl, 643, L119

\bibitem[{Carney} et~al.(1996){Carney}, {Laird}, {Latham} \&
  {Aguilar}]{1996AJ....112..668C}
{Carney} B.~W., {Laird} J.~B., {Latham} D.~W., {Aguilar} L.~A., 1996, \aj, 112,
  668

\bibitem[{G{\"a}nsicke} et~al.(2009){G{\"a}nsicke}, {Dillon}, {Southworth}
  et~al.]{2009MNRAS.397.2170G}
{G{\"a}nsicke} B.~T., {Dillon} M., {Southworth} J., et~al., 2009, \mnras, 397,
  2170

\bibitem[{Gianninas} et~al.(2006){Gianninas}, {Bergeron} \&
  {Fontaine}]{2006AJ....132..831G}
{Gianninas} A., {Bergeron} P., {Fontaine} G., 2006, \aj, 132, 831

\bibitem[{Horne} et~al.(1994){Horne}, {Marsh}, {Cheng}, {Hubeny} \&
  {Lanz}]{1994ApJ...426..294H}
{Horne} K., {Marsh} T.~R., {Cheng} F.~H., {Hubeny} I., {Lanz} T., 1994, \apj,
  426, 294

\bibitem[{Howell} et~al.(1997){Howell}, {Rappaport} \&
  {Politano}]{1997MNRAS.287..929H}
{Howell} S.~B., {Rappaport} S., {Politano} M., 1997, \mnras, 287, 929

\bibitem[{Hubeny}(1988)]{1988CoPhC..52..103H}
{Hubeny} I., 1988, Computer Physics Communications, 52, 103

\bibitem[{Hubeny} \& {Lanz}(1995)]{1995ApJ...439..875H}
{Hubeny} I., {Lanz} T., 1995, \apj, 439, 875

\bibitem[{Knigge}(2006)]{2006MNRAS.373..484K}
{Knigge} C., 2006, \mnras, 373, 484

\bibitem[{Knigge} et~al.(2011){Knigge}, {Baraffe} \&
  {Patterson}]{2011arXiv1102.2440K}
{Knigge} C., {Baraffe} I., {Patterson} J., 2011, ArXiv e-prints

\bibitem[{Kolb}(1993)]{1993A&A...271..149K}
{Kolb} U., 1993, \aap, 271, 149

\bibitem[{Littlefair} et~al.(2008){Littlefair}, {Dhillon}, {Marsh}
  et~al.]{2008MNRAS.388.1582L}
{Littlefair} S.~P., {Dhillon} V.~S., {Marsh} T.~R., et~al., 2008, \mnras, 388,
  1582

\bibitem[{Littlefair} et~al.(2007){Littlefair}, {Dhillon}, {Marsh},
  {G{\"a}nsicke}, {Baraffe} \& {Watson}]{2007MNRAS.381..827L}
{Littlefair} S.~P., {Dhillon} V.~S., {Marsh} T.~R., {G{\"a}nsicke} B.~T.,
  {Baraffe} I., {Watson} C.~A., 2007, \mnras, 381, 827

\bibitem[{Patterson} et~al.(2008){Patterson}, {Thorstensen} \&
  {Knigge}]{2008PASP..120..510P}
{Patterson} J., {Thorstensen} J.~R., {Knigge} C., 2008, \pasp, 120, 510

\bibitem[{Szkody} et~al.(2005){Szkody}, {Henden}, {Fraser}
  et~al.]{2005AJ....129.2386S}
{Szkody} P., {Henden} A., {Fraser} O.~J., et~al., 2005, \aj, 129, 2386

\bibitem[{Szkody} et~al.(2010){Szkody}, {Mukadam}, {G{\"a}nsicke}
  et~al.]{2010ApJ...710...64S}
{Szkody} P., {Mukadam} A., {G{\"a}nsicke} B.~T., et~al., 2010, \apj, 710, 64

\bibitem[{Townsley} \& {Bildsten}(2003)]{2003ApJ...596L.227T}
{Townsley} D.~M., {Bildsten} L., 2003, \apjl, 596, L227

\bibitem[{Townsley} \& {G{\"a}nsicke}(2009)]{2009ApJ...693.1007T}
{Townsley} D.~M., {G{\"a}nsicke} B.~T., 2009, \apj, 693, 1007

\end{thebibliography}
   
\bsp

\label{lastpage}

\end{document}